\documentclass[Crown,sagev,times]{sagej}

\usepackage{moreverb,url}
\usepackage{soul}
\usepackage{xcolor}

\definecolor{darkblue}{rgb}{0.0, 0.0, 0.5} 

\usepackage[hidelinks,colorlinks,bookmarksopen,bookmarksnumbered,citecolor=darkblue,urlcolor=darkblue,linkcolor=darkblue]{hyperref}

\usepackage{graphicx}

\newcommand\BibTeX{{\rmfamily B\kern-.05em \textsc{i\kern-.025em b}\kern-.08em
T\kern-.1667em\lower.7ex\hbox{E}\kern-.125emX}}

\def\volumeyear{2025}

\begin{document}

\runninghead{Jeremy J. Williams et al.}

\title{Integrating High Performance In-Memory Data Streaming and In-Situ Visualization in Hybrid MPI+OpenMP PIC MC Simulations Towards Exascale}


\author{Jeremy J. Williams\affilnum{1},
Stefan Costea\affilnum{2},
Daniel Medeiros\affilnum{1},
Jordy Trilaksono\affilnum{3},
Pratibha Hegde \affilnum{1}
David Tskhakaya\affilnum{4},
Leon Kos\affilnum{2},
Ales Podolnik\affilnum{4}, 
Jakub Hromadka\affilnum{4},
Kevin A. Huck\affilnum{5}, 
Allen D. Malony\affilnum{5}, 
Frank Jenko\affilnum{3}, 
Erwin Laure\affilnum{6} and
Stefano Markidis\affilnum{1}}


\affiliation{\affilnum{1}KTH Royal Institute of Technology, Stockholm, Sweden\\ 
\affilnum{2}Faculty of Mechanical Engineering, University of Ljubljana, Ljubljana, Slovenia\\
\affilnum{3}Max Planck Institute for Plasma Physics, Garching, Germany\\
\affilnum{4}Institute of Plasma Physics of the CAS, Prague, Czech Republic\\
\affilnum{5}University of Oregon, Eugene, Oregon, The United States of America\\
\affilnum{6}Max Planck Computing and Data Facility, Garching and Greifswald, Germany}

\corrauth{Jeremy J. Williams \\
Department of Computer Science, KTH Royal Institute of Technology, \\
Lindstedtsvägen 5, SE-100 44, Stockholm, Sweden \\ 
E-mail address: jjwil@kth.se}

\begin{abstract}
Efficient simulation of complex plasma dynamics is crucial for advancing fusion energy research. Particle-in-Cell (PIC) Monte Carlo (MC) simulations provide insights into plasma behavior, including turbulence and confinement, which are essential for optimizing fusion reactor performance. Transitioning to exascale simulations introduces significant challenges, with traditional file input/output (I/O) inefficiencies remaining a key bottleneck. This work advances BIT1, an electrostatic PIC MC code, by improving the particle mover with OpenMP task-based parallelism, integrating the openPMD streaming API, and enabling in-memory data streaming with the ADIOS2 Sustainable Staging Transport (SST) engine to enhance I/O performance, computational efficiency, and system storage utilization. We employ profiling tools such as gprof, perf, IPM and Darshan, which provide insights into computation, communication, and I/O operations. We implement time-dependent data checkpointing with the openPMD API enabling seamless data movement and in-situ visualization for real-time analysis without interrupting the simulation. We demonstrate improvements in simulation runtime, data accessibility and real-time insights by comparing traditional file I/O with the ADIOS2 BP4 and SST backends. The proposed hybrid BIT1 openPMD SST enhancement introduces a new paradigm for real-time scientific discovery in plasma simulations, enabling faster insights and more efficient use of exascale computing resources.
\end{abstract}


\maketitle

\section{Introduction}
Simulating complex physical phenomena at large scale is crucial for advancing fusion energy research. Particle-in-Cell (PIC) Monte Carlo (MC) simulations provide valuable insights into plasma dynamics, including turbulence, instabilities, and confinement properties, which are essential for optimizing fusion reactors. However, achieving efficient exascale simulations requires overcoming challenges in high-performance I/O, data streaming, and in-situ visualization for seamless data management and real-time analysis.

\begin{figure*}[!ht]
    \vspace{0cm} 
    \begin{center}
        \includegraphics[width=\linewidth]{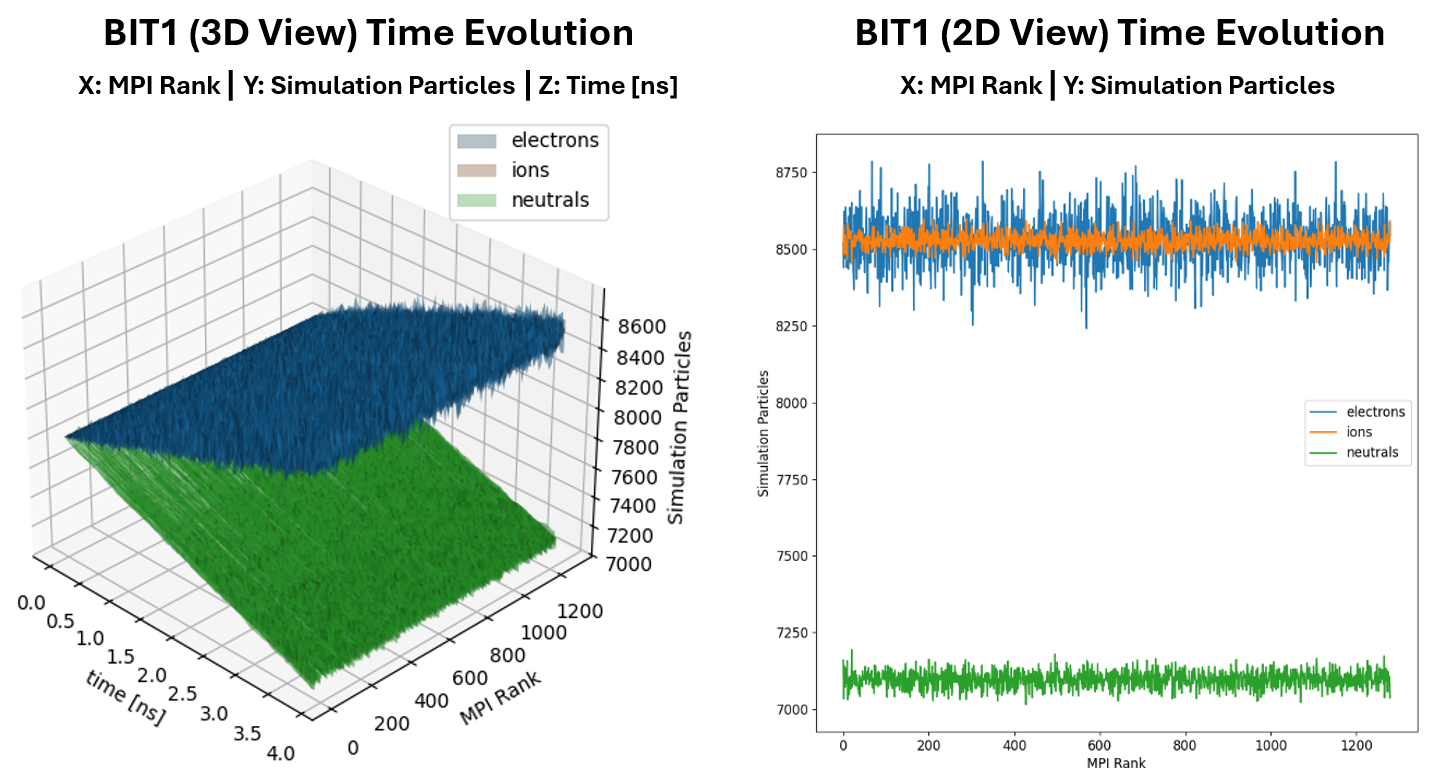}
        \caption{Displaying the time evolution (left) and the final state (right) of the total number of particles per MPI Rank (CPU Core) for each species in BIT1, using a standard Python script that reads simulation output data, for up to 10 nodes, after the simulation has completed.} \label{Post_Processing_BIT1}
        \vspace{-0.8cm} 
    \end{center}
\end{figure*}

PIC MC simulations using hybrid MPI+OpenMP parallelism optimize data throughput by leveraging MPI for distributed-memory communication and OpenMP for shared-memory parallelism. While this hybrid approach improves resource utilization and reduces synchronization overhead, traditional file I/O remains a major bottleneck. Frequent disk writes and redundant data dumps slow down computations, overload storage systems, and delay analysis, making it difficult to extract meaningful results efficiently. Conventional post-processing workflows in large-scale PIC MC simulations introduce further inefficiencies. As shown in Fig.~\ref{Post_Processing_BIT1}, a BIT1 user typically checks the results by displaying the time evolution or the final state of the total number of particles per CPU for each species. This is usually done using a standard Python script that processes large output files and generates plots after the simulation has completed. This approach delays insight generation, limits opportunities for real-time diagnostics, and is often followed by costly reruns if specific data were not recorded. As simulations scale, the overhead of writing massive datasets to disk and subsequently reading them back for offline analysis becomes increasingly unsustainable. These constraints highlight the growing need for integrated in-situ data processing and visualization techniques that provide immediate feedback, reduce I/O pressure, and streamline scientific workflows at exascale.

To address this, we introduce and integrate high-performance in-memory data streaming and in-situ visualization into a hybrid MPI+OpenMP version of BIT1 using openPMD and the ADIOS2 SST backend, focusing on streaming I/O performance, real-time checkpointing, and in-situ visualization. This approach builds on previous work~\cite{williams2024enabling,williams2024understanding}, in which BIT1’s traditional file I/O has transitioned to ADIOS2 streaming workflows coupled with OpenMP task-based parallelism, enabling fine-grained workload distribution to optimize data movement and computation for better scalability. The contributions of this work include:
\begin{itemize}
    \item We implement, for the first time, a hybrid MPI+OpenMP version of BIT1 integrated with openPMD to further enhance I/O performance, efficiency, and storage utilization. This approach addresses particle load imbalance through task-based parallelism and leverages the openPMD streaming API for both stream-based and file-based datasets. This development allows BIT1 to handle much larger simulations, reduce I/O bottlenecks, and scale efficiently across heterogeneous systems.
    \item We identify computationally intensive parts using an I/O adaptor for the openPMD interface with ADIOS2 SST, assessing BIT1’s performance on a single node. This analysis offers a systematic understanding of where the code spends the most time
    \item We employ profiling and monitoring techniques to understand the impact and performance of openPMD’s streaming API with ADIOS2 in scaling tests, comparing streaming-based workflows to traditional file I/O and the ADIOS2 BP4 engine when diagnostics are activated. These insights quantify the benefits of streaming I/O in practice. 
    \item We use a customized Python script with openPMD’s streaming API and ADIOS2 SST for real-time data checkpointing and visualization without interrupting the simulation, tailored for BIT1 output. This capability allows scientists and users of BIT1 to monitor and analyze simulations as they run, reducing time-to-discovery, improving fault tolerance with continuous checkpoints, and supporting interactive exploration of large-scale data.
\end{itemize}

The remainder of this paper is organized as follows. Section~2 provides background information on the openPMD Standard, openPMD-api \& ADIOS Version 2, and BIT1's advancements. Section~3 details our methodology and the experimental setup, including BIT1 modifications. Performance analysis \& visualization results are presented in Section~4. Related work is discussed in Section~5. Finally, Section~6 contains the discussion, conclusion, and future work.


\begin{figure*}[!ht]
   \vspace{0cm} 
    \begin{center}
        \includegraphics[width=\linewidth]{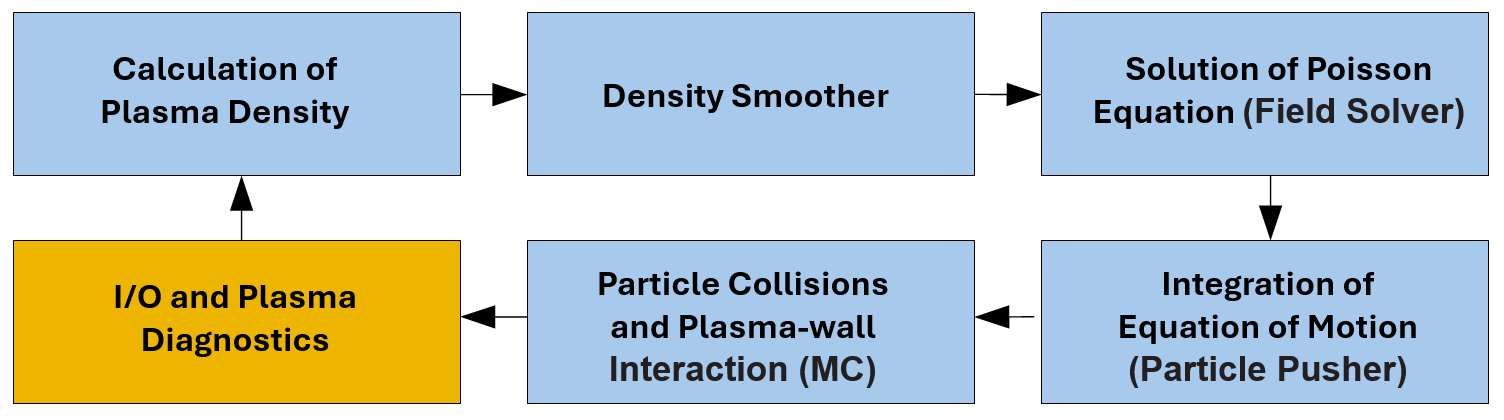}
        \caption{A simplified diagram representing BIT1 PIC MC workflow used to simulate the plasma edge~\cite{tskhakaya2004magnetised,tskhakaya2010pic}. After initialization, the PIC MC algorithm cycle repeats at each time step for diagnostic output and I/O performance investigations~\cite{williams2023leveraging}. In orange, we highlight the I/O and plasma diagnostics step activated to integrate high-performance in-memory data streaming and in-situ visualization in BIT1 using openPMD.} \label{BIT1_electrostatic_PIC_MC_code_workflow}
        \vspace{-0.4cm} 
    \end{center}
\end{figure*}

\section{Background}
The PIC method simulates plasma behavior by capturing the dynamics of charged/neutral particles in 1D/2D/3D spatial dimensions and resolving their 3D velocity distribution. It is particularly effective for studying plasma edge physics, where interactions with walls and particle collisions are crucial. PIC simulations often integrate MC techniques for accurate treatment of elastic and inelastic particle collisions, as well as plasma-wall interactions. BIT1 uses an explicit formulation of the PIC method and follows a well-defined workflow, as shown in Fig.~\ref{BIT1_electrostatic_PIC_MC_code_workflow}. BIT1’s PIC MC cycle involves five phases~\cite{tskhakaya2010pic,williams2023leveraging}: interpolating particle properties, solving the electric field, handling collisions, weighting forces to particles, and advancing particle trajectories. Alongside these five phases and outlined in Fig.~\ref{BIT1_electrostatic_PIC_MC_code_workflow}, the I/O and plasma diagnostics phase (in orange) occurs only for a specific number of time steps (for instance, only every 1,000 cycles), enabling key diagnostics and providing checkpointing and restart capabilities.

Berkeley Innsbruck Tbilisi 1D3V (BIT1) is a specialized 1D3V electrostatic PIC MC code~\cite{tskhakaya2002pic} and based on the \texttt{XPDP1} code, developed by Birdsall's team at the University of California, Berkeley (UC Berkeley) in the 1990s~\cite{verboncoeur1993simultaneous}. It consists of approximately 31,600 lines of code and is written entirely in the C programming language. At present, BIT1 does not depend on external numerical libraries but instead implements native solvers for the Poisson equation, particle mover, and smoother. BIT1~\cite{tskhakaya2007optimization,tskhakaya2010pic} is designed to study high-density plasma, impurity, and neutral transport in one-dimensional magnetic flux tubes representing the edge region of magnetically confined fusion devices. Despite its reduced dimensionality, BIT1 can resolve a wide range of kinetic effects \cite{tskhakaya2021compass,costea2021blobs}(e.g. non-Maxwellian distributions, non-local transport, atomic and molecular processes, etc.), enabling critical investigations into plasma-wall interactions, sheath formation, and impurity transport.

Although the original version of BIT1 features reliable serial I/O (i.e. small and compressed binary files for fast dump/checkpointing), the growing scale and complexity of simulations have highlighted the limitations of this approach. As data volumes increased, serial I/O became a performance bottleneck, with an increased risk of file corruption and significant delays in data writing. These challenges prompted the development and integration of parallel I/O methods~\cite{williams2024enabling} to ensure reliable and scalable performance for large-scale simulations.

BIT1 is driven by a small input file (1–3 kB), ensuring efficient initialization across all processes. Its output framework consists of key parameters that control data diagnostics, state preservation, and time-dependent averaging of plasma profiles and velocity distributions. Two critical input parameters guide the output analysis~\cite{williams2024enabling,williams2024understanding}:

\begin{itemize}
\item \textbf{mvflag}: Activates and enables time-dependent diagnostics of plasma profiles and particle angular, velocity, and energy distribution functions. If greater than zero, it specifies the number of time steps over which the diagnostics are averaged.
\item \textbf{mvStep}: Specifies the number of time steps between each time-dependent diagnostic output.
\end{itemize}

Unlike PIC codes such as Smilei~\cite{derouillat2018smilei} and Warp-X~\cite{vay2018warp}, which are optimized for high-performance plasma simulations but not for high-density plasma, BIT1 is specifically tailored for plasma edge modelling, especially the Scrape-Off Layer (SOL) in fusion devices. BIT1 addresses SOL challenges such as kinetic effects, plasma-wall interactions, and sheath physics by incorporating Direct Simulation Monte Carlo (DSMC) collision operators~\cite{tskhakaya2023implementation} and is capable of modeling next-generation fusion devices with an extremely high-density plasma edge. As the first massively parallel full-orbit Debye-scale PIC code applied to SOL simulations~\cite{tskhakaya2012recent}, BIT1 uses the ``natural sorting" method~\cite{tskhakaya2007optimization} to improve computational efficiency, accelerating collision operators by up to five times. However, ``natural sorting" requires more memory per grid cell, creating a bottleneck when offloading to GPUs~\cite{williams2025accelerating}, and regions with high particle concentrations may cause workload imbalances~\cite{williams2023leveraging}, requiring adaptive parallel strategies. Given this, BIT1 is evolving toward a hybrid computational model with MPI and OpenMP to ensure portability across supercomputing platforms. BIT1 uses MPI-based domain decomposition for parallel execution and OpenMP for shared-memory parallelism, reducing MPI communication overhead and improving scalability on multi-core processors\cite{williams2024optimizing}.

\begin{figure*}[!ht]
    \vspace{0cm} 
    \begin{center}
        \includegraphics[width=\linewidth]{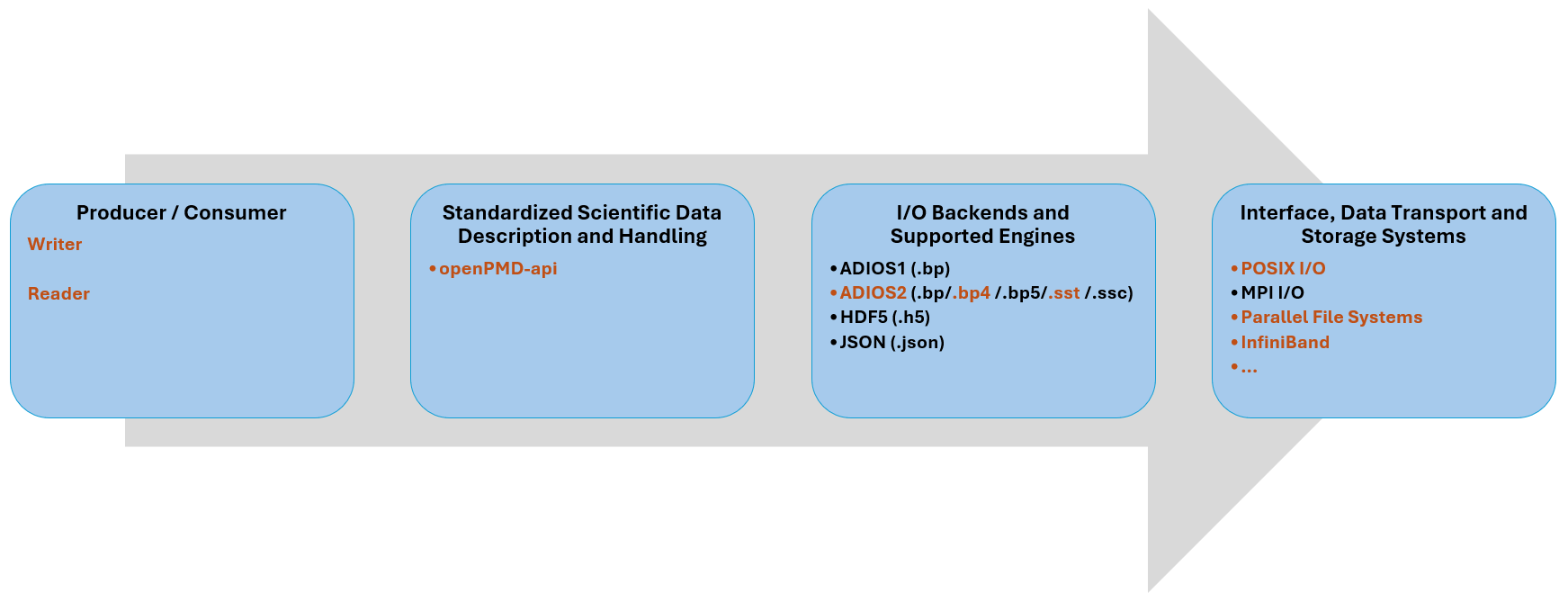}
        \caption{An Open Software Stack for Scientific I/O~\cite{openPMDapi,openPMDstandard,poeschel2021transitioning} integrates high-performance in-memory data streaming and in-situ visualization in BIT1 using openPMD. In orange, we highlight the BP4 engine, which is aggressively optimized for I/O efficiency at large scale by reducing metadata and aggregating data, and the SST engine, which optimizes communication and manages data transfers with minimal overhead, making it ideal for exascale systems.} \label{openPMD_Software_Stack_Scientific_IO}
        \vspace{-0.4cm} 
    \end{center}
\end{figure*}

\subsection{openPMD Standard \& openPMD-api}
The openPMD (Open Standard for Particle-Mesh Data) standard~\cite{openPMDstandard} enables the portable exchange of particle and mesh data, supporting formats such as HDF5, ADIOS1, ADIOS2, and JSON in serial and MPI workflows. The openPMD-api~\cite{openPMDapi} facilitates scientific I/O, as shown in Fig.~\ref{openPMD_Software_Stack_Scientific_IO} across these formats, with records representing physical quantities and iterations over time. In openPMD, a record is a physical quantity of arbitrary dimensionality (rank), potentially consisting of multiple components (e.g., scalars, vectors, tensors), and these records share common properties, such as describing electric fields, density fields, or particle attributes. Records may be structured as meshes (n-dimensional arrays) or unstructured, as in the case of particle species stored in 1D arrays, where each row corresponds to a particle. Updates to the values of meshes or particle species are called iterations, which represent the temporal evolution of records. A collection of such iterations forms a series, and the openPMD-api supports this concept through a variety of backends and encoding strategies.

\subsection{ADIOS Version 2}
The ADIOS2 (Adaptable Input Output System version 2) is an open-source, scalable parallel I/O framework designed for efficient data transfer on supercomputers, with support for C, C++, Fortran, and Python. A key feature of ADIOS2 is its use of virtual I/O engines, which define how data is read and written and provide configurable settings to optimize performance for various workloads. The available engines include the Binary Packed formats versions 5 (BP5), 4 (BP4) and 3 (BP3), the Hierarchical Data Format version 5 (HDF5), Sustainable Staging Transport (SST), Strong Staging Coupler (SSC), DataMan and Inline. Previous work~\cite{openpmdbestpractice,williams2024enabling,williams2024understanding} focused on the BP4 engine, which aggressively optimizes simulations for I/O efficiency at large scale through large per process buffers and minimal write operations, enabling high throughput on parallel file systems, whereas the BP5 engine introduces certain compromises to achieve tighter control of host memory by using a linked list of equally sized buffer chunks (up to 2GB each, configurable via the ADIOS2 ``BufferChunkSize" parameter) and requires careful configuration of aggregators, subfiles and buffer flushing. While this improves memory predictability and mitigates out of memory risks, it generally delivers lower I/O throughput than BP4. In this work, we focus on the SST engine to enable streaming support in BIT1. SST is optimized for communication and large-scale data transfers with minimal overhead, making it particularly well suited for exascale systems, and it integrates seamlessly with openPMD to support efficient data exchange in MPI parallel workflows.


\section{Methodology \& Experimental Setup}
This work integrates a hybrid MPI+OpenMP version of BIT1 with the openPMD standard and the ADIOS2 SST backend to enhance I/O efficiency, support parallel streaming workflows, minimize data storage, and enable real-time data checkpointing and in-situ visualization, while detailing the specific modifications implemented in BIT1.

\subsection{Hybrid BIT1 openPMD \& ADIOS2 SST Implementation}
 \noindent \textbf{OpenMP Tasks Particle Mover Parallelization.} OpenMP is a widely used programming model for shared-memory parallelism in HPC, supported by major compilers such as GCC and LLVM, ensuring broad accessibility.
 
Williams et al.~\cite{bit1openmptasksmover,williams2024optimizing} presented an OpenMP implementation of the particle mover's core function. In this function, the arrays \texttt{x[species][cell][particle]} and \texttt{vx[species][cell][particle]} store particle positions and velocities, while \texttt{nsp}, \texttt{nc}, and \texttt{np[species][cell]} represent the number of plasma species, grid cells, and particles per species per cell \cite{tskhakaya2007optimization}.
The OpenMP \texttt{taskloop} construct is employed to parallelize the outer loop, which has a limited number of iterations, ensuring efficient load balancing on multicore CPUs. The \texttt{\#pragma omp parallel} directive initializes a parallel region with shared variables, while \texttt{isp} and \texttt{i} remain private. The \texttt{firstprivate} clause ensures that \texttt{nsp} and \texttt{nc} are private and initialized for each thread.
Within the parallel region, the \texttt{single} construct ensures that initialization is executed by a single thread. The \texttt{taskloop grainsize(500) nogroup} pragma dynamically distributes iterations, optimizing task granularity. Additionally, the \texttt{simd} pragma within the innermost loop enhances SIMD vectorization, improving the efficiency of particle movement calculations.

\noindent \textbf{openPMD \& openPMD-api Integration.} BIT1 is a 1D3V PIC MC code in C, simulating particles in one dimension with three velocity components. The openPMD-api parallel I/O library is integrated to enhance I/O functionality. The integration of openPMD and ADIOS2 in BIT1 for streaming and file output involves several key steps. First, the MPI environment is initialized, determining the number of processors and their ranks to ensure proper communication. The input file is then read, and necessary initializations are performed, including memory allocations and calculations for profiles and field values. If streaming is enabled, a series is created using the ADIOS2 SST engine with configurations specifying parameters such as transport type (e.g. WAN), buffer growth factor (e.g. 2.0), and queue policies (e.g. QueueLimit=2, QueueFullPolicy=Discard). During the main iteration loop, particle movements are simulated, boundary conditions are updated, and various calculations are performed based on specific flags. At defined intervals, data is written to ADIOS2 files in either BP4 or SST format. Time-dependent diagnostics, including system states and particle history, are saved, and when streaming with the SST engine is activated, data is written to memory instead of disk. Once the loop completes, the results are stored, resources are cleaned up, and the MPI environment is finalized. 

\begin{figure*}[!ht]
    \vspace{0cm} 
    \begin{center}
        \includegraphics[width=0.8\linewidth]{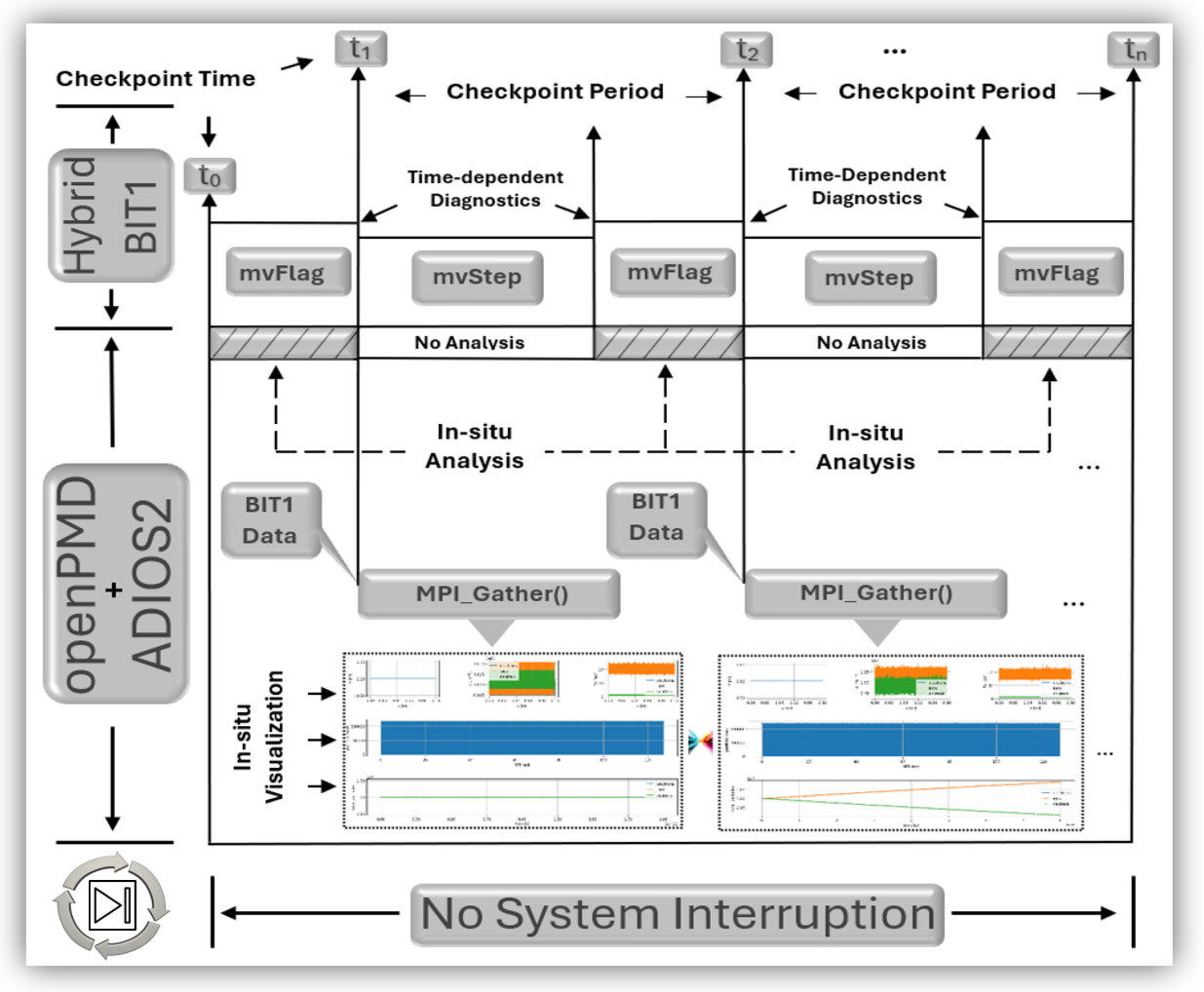}
        \caption{Hybrid BIT1 I/O and Insight Workflow using openPMD and ADIOS2 Backends.} 
        \label{Streaming_and_In-Situ_Visualization}
    \end{center}
    \vspace{-0.4cm} 
\end{figure*}

\begin{figure*}[!ht]
    \vspace{0cm} 
    \begin{center}
        \includegraphics[width=\linewidth]{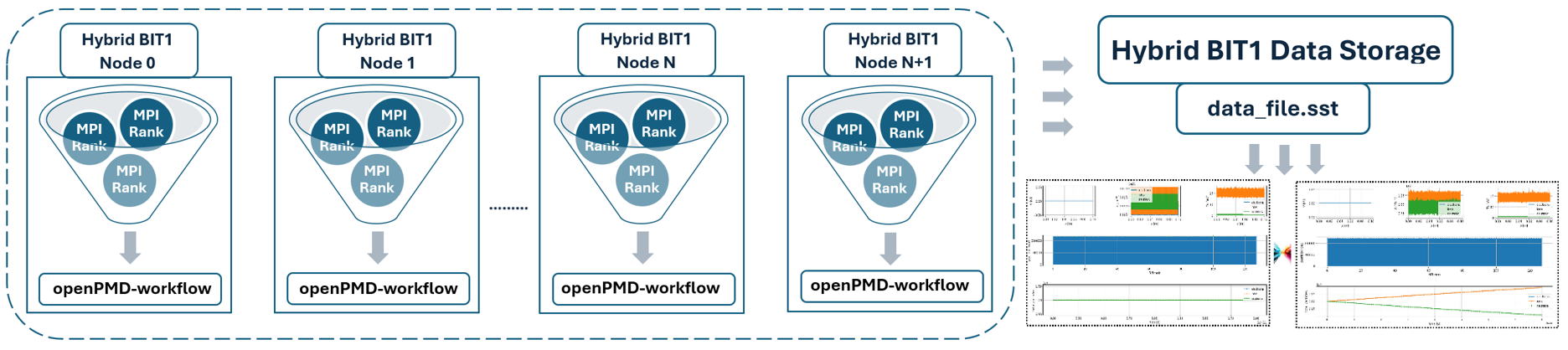}
        \caption{Hybrid BIT1 openPMD SST Workflow and In-Situ Visualization Without Interruption the Simulation.} 
        \label{Hybrid_BIT1_I/O_Workflow_using_ADIOS2_SST}
    \end{center}
    \vspace{-0.4cm} 
\end{figure*}

\noindent \textbf{openPMD \& ADIOS2 SST Backend Activation.} In \texttt{bit1.cpp}~\cite{bit1openpmdsst}, written in C++ to enable and activate streaming, the BIT1 executable must be run with the \texttt{--streaming} parameter, and the input file should include \texttt{origdmp=2} (0=serial I/O ASCII, 1=serial I/O binary) to enable parallel writing with ADIOS2 and \texttt{mvFlag > 0} (integer, number of steps to average from) to prepare the streaming data. The \texttt{Series} object is configured with ADIOS2 settings, defining parameters such as buffer growth, transport type, and queue limits. Streaming occurs at a frequency of \texttt{mvFlag+mvStep}, with profiles averaged over \texttt{mvFlag} steps, and system states saved at specific intervals. The iteration is finalized by closing the stream, clearing buffers, and advancing the iteration number. Two types of single-aggregator consumers can capture the stream in our setup: an openPMD pipe for checkpointing (\texttt{openpmd-pipe --infile$=$input\_file.sst --outfile$=$input\_file.bp4}) and in-situ visualization for real-time analysis (\texttt{python3 in-situ-vis.py input\_file.sst}). Both can run simultaneously due to the queue limit setting, and visualizations at every checkpoint can also be generated from time-dependent checkpoint data stored on disk (\texttt{python3 in-situ-vis.py input\_file.bp4}). Fig.~\ref{Streaming_and_In-Situ_Visualization} and~\ref{Hybrid_BIT1_I/O_Workflow_using_ADIOS2_SST} displays the Hybrid BIT1 I/O Workflow with openPMD using the ADIOS2 SST Backend, output file \texttt{data\_file.sst} supports real-time capability ensuring the analysis and visualization occurs promptly at every checkpoint without causing any interruption to the simulation.

\begin{figure*}[!ht]
    \vspace{0cm} 
    \begin{center}
        \includegraphics[width=\linewidth]{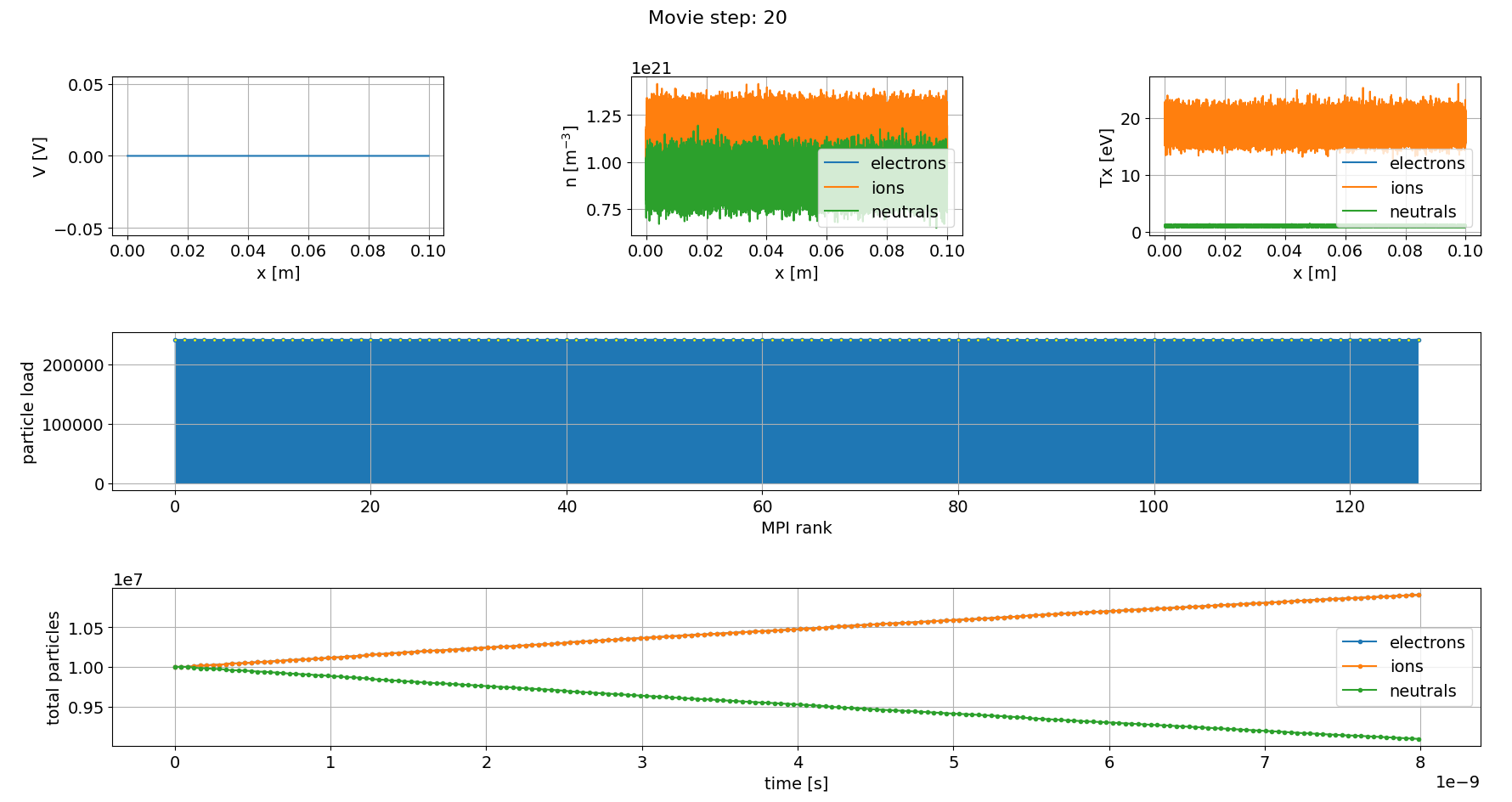}
        \caption{Hybrid BIT1 In-Situ Visualization Output at the End of the Simulation, as in Fig.~\ref{Streaming_and_In-Situ_Visualization},~\ref{Hybrid_BIT1_I/O_Workflow_using_ADIOS2_SST} and~\ref{BIT1_openPMD_In-situ_Visualizations_Python}.} 
        \label{Analysis_and_In-Situ_Visualization_Output}
    \end{center}
    \vspace{-0.8cm} 
\end{figure*}

\subsection{HPC Profiling \& Monitoring Tools}
In this work, we aim to understand the impact of the hybrid version of BIT1 with the openPMD and ADIOS2 SST backend, determining its associated performance characteristics. To achieve this, we employ a suite of sophisticated profiling and monitoring tools. Specifically, we utilize:
\begin{itemize}
    \item \textbf{gprof}, an open-source profiling tool that collects execution time data and identifies frequently used functions, generating separate outputs for each MPI process, which are then consolidated into one report.
    \item \textbf{perf}, a low-level profiler designed to gather key performance metrics, such as total execution time, instructions executed, cache behavior, and CPU cycles.
    \item \textbf{IPM} (Integrated Performance Monitoring), a performance profiling tool that captures the computation and communication activities of BIT1, providing detailed reports on MPI calls and buffer sizes.
    \item \textbf{Darshan}, a performance monitoring tool specifically designed for analyzing serial and parallel I/O workloads.
\end{itemize} 
\subsection{Use Case \& Experimental Environment}
We aim to improve both scalability and I/O efficiency in hybrid BIT1 by utilizing the openPMD and ADIOS2 backends. The test case involves an initial uniformly distributed hot plasma interacting with a uniformly distributed monoatomic gas, which ultimately leads to the ionization of the gas and the cooling of the plasma. The initial conditions consist of three particle species: electrons, single-ionized deuterium ions (D\(^+\)), and deuterium neutrals (D), all uniformly distributed inside the system, with electron and ion initial densities of \(10^{21}\, \text{m}^{-3}\) and initial temperatures of 20 eV for both charged species, while the neutral species has a density of \(10^{21}\, \text{m}^{-3}\) and an initial temperature of 1 eV. The simulation assumes no electric or magnetic fields. Periodic boundary conditions are applied, meaning particles crossing a boundary are reinjected from the opposite boundary with the same velocities, resulting in no plasma-wall interaction. The computational mesh consists of 100K cells, each with 100 particles per species, yielding a system size of 1 meter and a cell size of \(10\, \mu\text{m}\) (micrometres), resulting in a total of 30M particles. The simulation runs for 200K time steps, each with a time step of \(4 \times 10^{-14}\, \text{s}\), simulating a total of 8 ns of plasma evolution. Particle collisions considered in the simulation include electron-neutral elastic, excitation, and ionization processes. Unless otherwise noted, the simulation runs for up to 200K time steps. The number of time steps/cycles was chosen for convenience for profiling purposes only, and it does not relate to a reached steady state or a certain maturity of the simulation. It is important to note that this test excludes the field solver and smoother phases, as in~\cite{williams2024enabling,williams2024optimizing,williams2024understanding,williams2025accelerating}, since this use case assumes no electric field, and the field solver and smoother functions are never called in this particular use case.

We simulate and evaluate the impact of integrating openPMD with hybrid BIT1 using the ADIOS2 SST backend on the following three distinct systems:
\begin{itemize}
\item \textbf{Dardel}, an HPE Cray EX supercomputer, features a CPU partition with 1270 compute nodes. Each node is equipped with two AMD EPYC™ Zen2 2.25 GHz 64-core processors, 256 GB DRAM, and interconnected using an HPE Slingshot network with the Dragonfly topology, amounting to 200 GiB/s Bandwidth. In terms of storage, Dardel has a Lustre File System (LFS) with 12 PB in capacity and 48 OSTs. The OS is SUSE Linux Enterprise Server 15 SP3, and all applications were compiled with GCC 11.2, openPMD 0.15.2, ADIOS2 2.10.0 (with Blosc and bzip2 compression enabled), and Cray MPICH 8.1 as the MPI flavor for intra-node communication.
\item \textbf{Discoverer}, a petascale EuroHPC supercomputer, features a CPU partition with 1128 compute nodes. Each node is equipped with two AMD EPYC 7H12 64-Core processors, 256 GB DDR4 SDRAM (on regular nodes), 1TB DDR4 SDRAM (on fat nodes), interconnected using Ethernet Controller I350 with 10 GiB/s Bandwidth and Mellanox ConnectX-6 InfiniBand with the Dragonfly+ topology, amounting to 200 GiB/s Bandwidth. For storage, \textbf{Discoverer} has a Network File System (over Ethernet) with 4.4 TB, and LFS with 2.1 PB in capacity and 4 Object Storage Targets (OSTs). The operating system (OS) is Red Hat Enterprise Linux release 8.4, and all the applications were compiled with GCC 11.4.0 and MPI library, MPICH 4.1.2 for intra-node communication.
\item \textbf{Vega}, a petascale EuroHPC supercomputer, features a CPU partition with 960 compute nodes. Each node is equipped with two AMD EPYC 7H12 64-Core processors, 256 GB DDR4 SDRAM (80\%/nodes), 1TB DDR4 SDRAM (20\%/nodes), interconnected using Mellanox ConnectX-6 InfiniBand HDR100 with a Dragonfly+ topology, amounting up to 500 GiB/s Bandwidth. For storage, \textbf{Vega} has a Ceph File System (CephFS) with 23 PB, and LFS with 1 PB in capacity and 80 OSTs. The OS is Red Hat Enterprise Linux 8, and all applications were compiled with GCC 12.3.0 and the MPI library, OpenMPI 4.1.2.1 for intra-node communication.
\end{itemize}

\subsection{Hybrid BIT1 openPMD Streaming I/O \& Insight Workflow}
As presented by Williams et al.~\cite{williams2023leveraging,williams2024optimizing}, hybrid BIT1 performs serial I/O operations throughout each simulation. Similar to the process described in~\cite{poeschel2021transitioning}, this work enables the ADIOS2 SST backend in hybrid BIT1 for streaming support. Unlike BP4, which writes data to files, SST enables direct data streaming between producers and consumers, reducing I/O overhead and improving scalability in large-scale systems. SST supports full \texttt{MxN} data distribution, allowing flexible reader and writer ranks, and enabling multiple reader cohorts to access data simultaneously. 

As shown in Fig.~\ref{Streaming_and_In-Situ_Visualization},~\ref{Analysis_and_In-Situ_Visualization_Output},~\ref{Hybrid_BIT1_I/O_Workflow_using_ADIOS2_SST} and presented in~\cite{williams2024enabling,williams2024understanding}, integrated into hybrid BIT1’s openPMD workflow, the simulation uses \texttt{mvFlag} to activate time-dependent diagnostics of plasma profiles and particle distributions, averaging over a specified number of time steps when \texttt{mvFlag > 0}, and \texttt{mvStep} to define the time step interval for diagnostics. SST utilizes RDMA network interconnects in HPC environments while maintaining compatibility with standard networking protocols, reducing storage bottlenecks, and enhancing I/O efficiency. When the ADIOS2 SST backend is used in hybrid BIT1’s openPMD workflow, it enables real-time data streaming across MPI ranks, supporting in situ analysis and large-scale data handling with real-time visualizations without blocking the simulation.

\section{Performance Results, Analysis \& Visualization}
In this work, we evaluate the impact and performance of integrating openPMD with hybrid BIT1 using the ADIOS2 SST backend. 

\begin{figure}[!ht]
    \vspace{0cm} 
    \begin{center}
        \includegraphics[width=\linewidth]{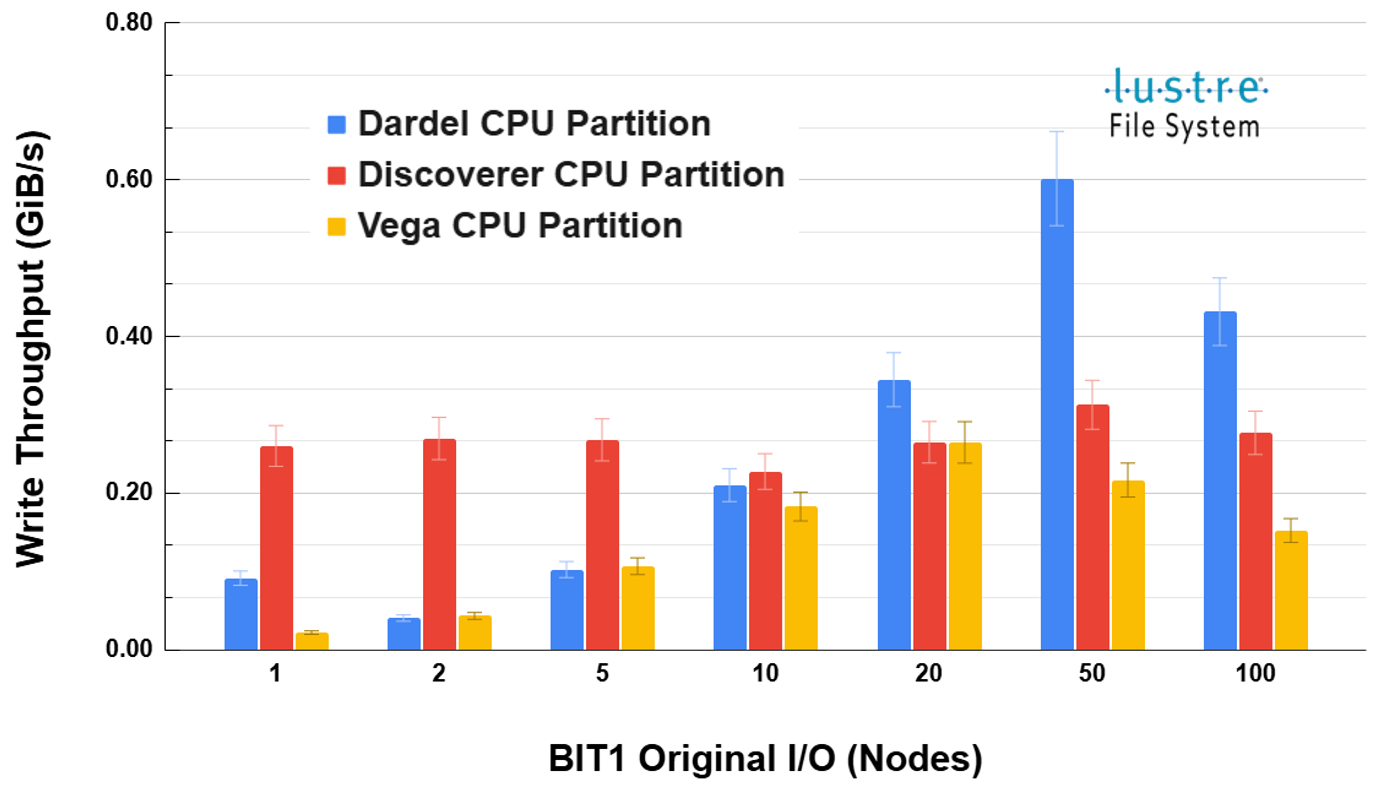}
        \caption{BIT1 Original File I/O Write Throughput with Variability ($>$10 runs), on \emph{Discoverer}, \emph{Dardel} and \emph{Vega} CPU Partitions, up to 100 Nodes (12,800 MPI processes), measured in GiB/s~\cite{williams2024enabling,williams2024understanding}.} \label{darshan_BIT1_Original_IO}
    \end{center}
    \vspace{-0.4cm} 
\end{figure}

\subsection{BIT1 Original Instrumentation \& I/O Monitoring} 
As previously investigated by Williams et al.~\cite{williams2023leveraging,williams2024understandingp2} and shown in Fig.~\ref{darshan_BIT1_Original_IO}, \texttt{Darshan}, a performance tool for analyzing I/O workloads, was used to assess BIT1's Original I/O write throughput (GiB/s) from its output logs across three supercomputing systems equipped with a LFS: \texttt{Dardel}, \texttt{Discoverer}, and \texttt{Vega}. Each data point represents the mean of more than ten runs, with error bars capturing the natural variability of I/O behavior. Since the Original BIT1 writes one file per MPI rank (and additional files for diagnostic results), the number of concurrent writers scales with node count, which in turn stresses the underlying file system differently on each machine. \texttt{Discoverer} exhibits a non-linear trend, with an initial throughput increase followed by a decline and a subsequent minor recovery, likely due to its limited 4 OSTs within a 2.1 PB LFS, which constrain parallel scalability. \texttt{Vega} shows a generally increasing trend but with noticeable variability at higher node counts, reflecting contention effects in its Lustre configuration. In contrast, \texttt{Dardel} achieves the highest peak throughput of 0.60 GiB/s at 50 nodes, enabled by its well-balanced architecture, including a 12 PB LFS with 48 OSTs and an HPE Slingshot interconnect (200 GiB/s), which efficiently handles parallel writes. Although throughput declines at 2 and 100 nodes, these variations fall within the normal range expected on Lustre systems. This behavior occurs because many MPI ranks attempt to write files simultaneously, placing additional pressure on the file system's metadata handling and potentially reducing overall performance. Among the three platforms, \texttt{Dardel} delivers the most stable and highest I/O performance, indicating its suitability for large-scale, I/O-intensive workloads. Based on these results, we recommend prioritizing further analysis and development efforts on the \texttt{Dardel} CPU LFS platform to maximize I/O efficiency.

\begin{figure}[!ht]
    \vspace{0cm} 
    \begin{center}
        \includegraphics[width=\linewidth]{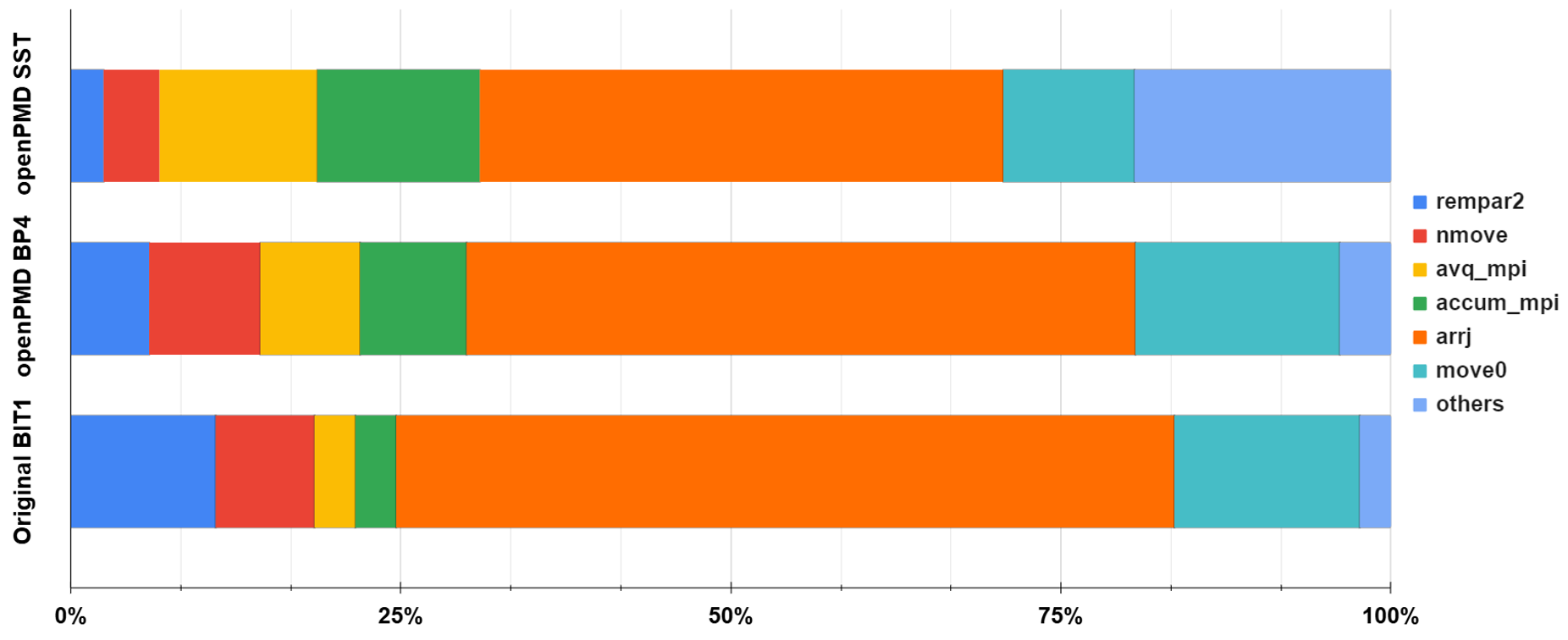}
        \caption{Percentage breakdown of the Ionization case functions on \emph{Dardel}, highlighting where most of the execution time is spent for the openPMD SST, openPMD BP4, and Original BIT1~\cite{williams2023leveraging,williams2024understanding} versions. The \texttt{arrj} sorting function (orange) was the most time-consuming, with its share of execution time reduced from 75.5\% in Original BIT1 to 65.5\% with openPMD BP4 and 35.5\% with openPMD SST. The \texttt{gprof} tool was used for profiling.} \label{function_breakdown}
     \end{center}
     \vspace{-0.4cm} 
\end{figure}

\subsection{BIT1 openPMD SST Performance \& I/O Costs Per Process}
We begin by utilizing \texttt{gprof}, an open-source profiling tool, to analyze execution time and identify the most frequently used functions across MPI processes. The consolidated \texttt{gprof} report provides a detailed performance analysis of the Original BIT1 with and without openPMD and the ADIOS2 backends.

As previously discovered by Williams et al.~\cite{williams2023leveraging,williams2024understandingp2}, Fig.~\ref{function_breakdown} shows how the most time consuming functions perform during the ``BIT1 openPMD SST", simulation compared to the ``BIT1 openPMD BP4" and ``Original BIT1" simulations. In the ``Original BIT1", \texttt{arrj} (a sorting function) dominates at 75.5\%, reducing to 65.5\% in BP4 and 35.5\% in SST, highlighting improved data handling. \texttt{move0} (a particle mover function) decreases from 18\% (Original) to 9.2\% (SST), while \texttt{rempar2} (a function used to remove particles) drops from 14\% to 2\%, indicating better parallelization. \texttt{nmove} (a particle mover function for the D neutral species) initially increases in BP4 but falls to 3.9\% in SST. \texttt{avq\_mpi} (a function used to compute profiles) and \texttt{accum\_mpi} (a function used to compute smoothed profiles) rise with BP4 and SST, reflecting enhanced MPI communication. The \texttt{others} category grows in SST, suggesting redistributed computation. Overall, the ``BIT1 openPMD SST" simulation spends the least time on the most time-consuming functions, significantly reducing the cost of \texttt{arrj} and \texttt{move0} compared to the ``BIT1 openPMD BP4" and ``Original BIT1" simulations, redistributing work across functions for better parallelization, more efficient use of computational resources, and improved overall simulation performance.

\begin{figure}[!ht]
    \vspace{0cm} 
    \begin{center}
        \includegraphics[width=\linewidth]{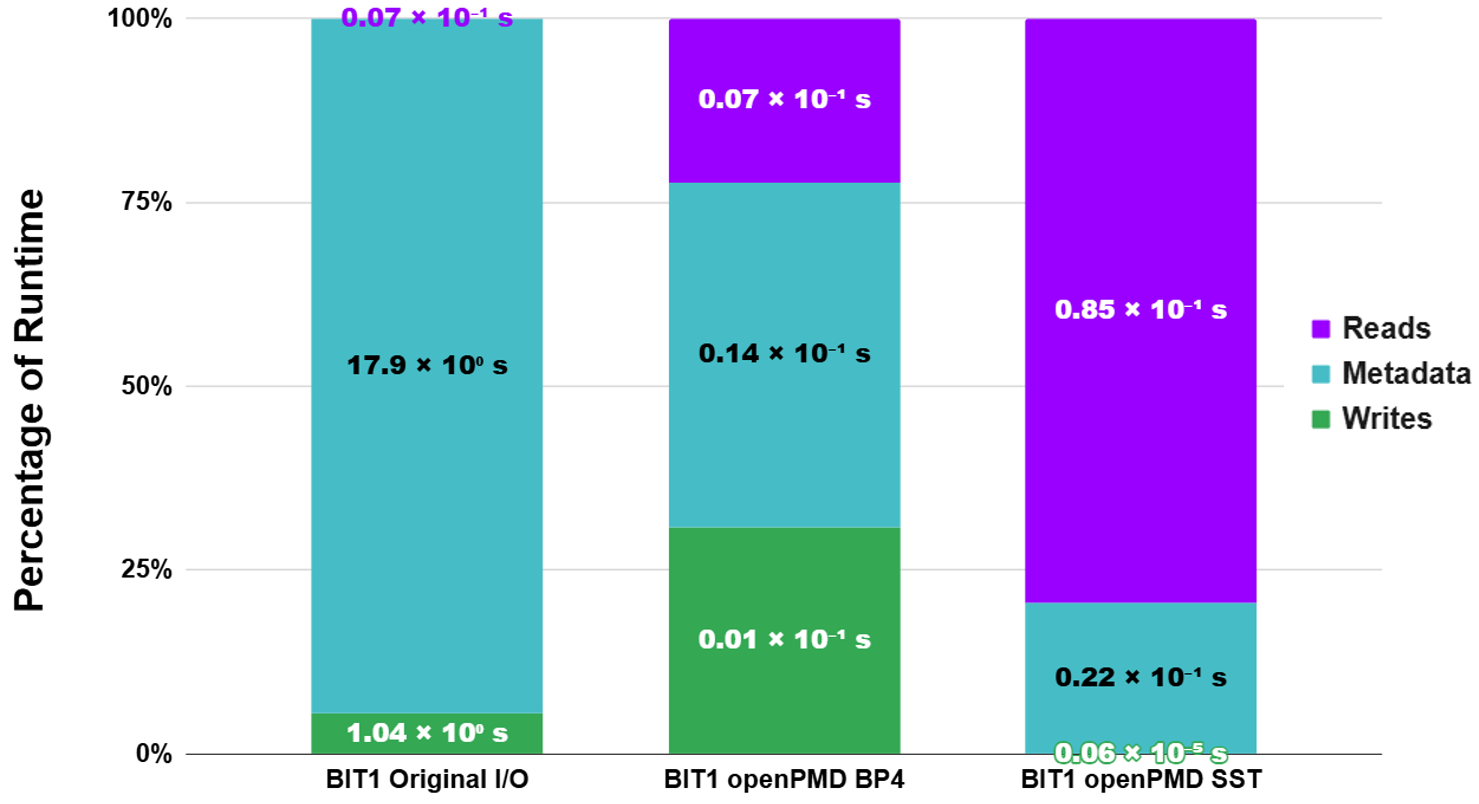}
        \caption{BIT1 Average I/O Cost Per Process for Reads, Metadata, and Writes on Dardel using 100 nodes, normalized.} 
        \label{BIT1_Runtime_Percentage_Time_Spent_100_nodes}
    \end{center}
    \vspace{-0.4cm} 
\end{figure}

Next, we utilized \texttt{Darshan}, a performance monitoring tool tailored for analyzing I/O workloads. As discovered by Williams et al.~\cite{williams2023leveraging,williams2024optimizing,williams2024understanding}, the peak I/O write throughput depends on the problem size, but once this peak is reached, performance degrades due to increased metadata writing costs in large runs. To understand the benefits of using openPMD with ADIOS2 backends, we investigate the time spent in I/O functions on 100 nodes, comparing BIT1's original I/O method with openPMD using BP4 and SST. Fig.~\ref{BIT1_Runtime_Percentage_Time_Spent_100_nodes} shows the normalized results of average I/O reads, writes, and metadata costs per process on 100 nodes.

Analyzing the results reveals that the integration of openPMD with ADIOS2 backends, particularly employing BP4 and SST, has yielded remarkable enhancements in BIT1. The use of BP4 has led to a dramatic reduction in metadata overhead, from 17.87s ($17.9 \times 10^{0}$ s) in the Original I/O to just 0.014s ($0.14 \times 10^{-1}$ s), a 99.92\% improvement. Write times also improved significantly, dropping from 1.04s ($1.04 \times 10^{0}$ s) in the Original I/O to 0.009s ($0.01 \times 10^{-1}$ s) with BP4. SST, while eliminating write times entirely at 0.0000006s ($0.06 \times 10^{-5}$ s), resulted in an increased read time of 0.085s ($0.85 \times 10^{1}$ s), likely due to its memory-streaming approach. BP4 achieves these improvements by consolidating data into a single file per node~\cite{williams2024enabling,williams2024understanding}, reducing metadata and write times, while SST streams data through memory, improving write performance but increasing memory usage and read time. Overall, BP4 provides the best I/O performance by balancing metadata reduction and efficient writes when running BIT1 openPMD BP4 simulations.

\subsection{Hybrid BIT1 openPMD SST Performance \& Scalability}
Previously presented by Williams et al.~\cite{williams2024optimizing}, the hybrid MPI+OpenMP version of BIT1 reduces total simulation and mover function time by leveraging OpenMP threads for concurrent execution. We extend this work by integrating openPMD with the ADIOS2 SST backend, using \texttt{IPM} to analyze MPI communication and load balancing from small-scale runs on a single node to large-scale runs on up to 100 nodes.

\begin{figure} [!ht]
    \vspace{0cm} 
    \begin{center}
       \includegraphics[width=\linewidth]{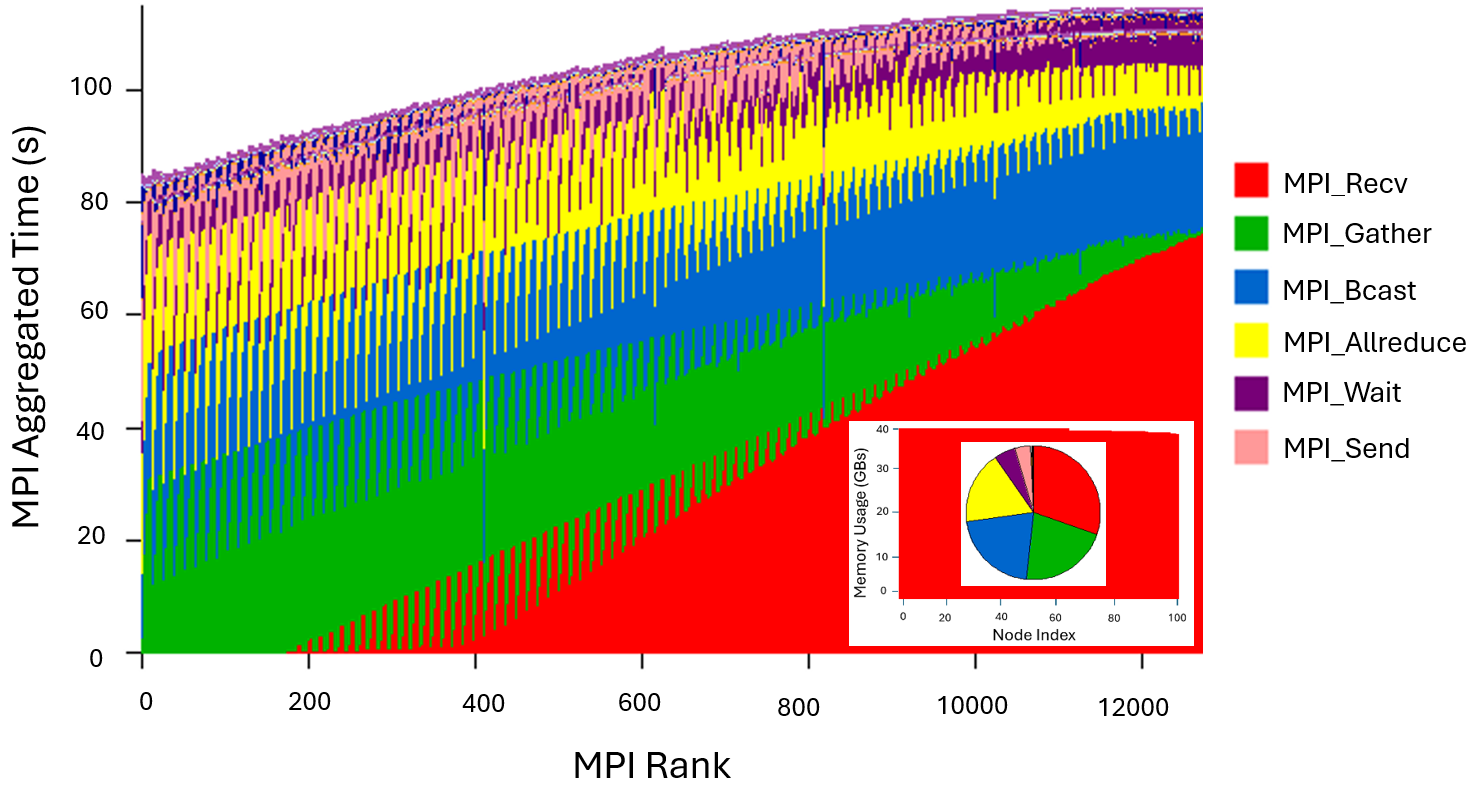} 
       \caption{MPI aggregated communication time for the Hybrid BIT1 openPMD SST simulation on \emph{Dardel}  using 100 nodes, totaling 12,800 MPI processes.} \label{MPI_Communication_and_Memory_Usage_SST}
    \end{center}
    \vspace{-0.4cm} 
\end{figure} 

As shown in the \texttt{IPM} profile ``index.html" output file, the hybrid BIT1 openPMD SST simulation on 100 nodes (12,800 cores) completed in 127.21 seconds, with 81.39\% of the execution time dedicated to MPI communication. The simulation achieved approximately 2.70 billion floating-point operations per second (2.7 GFLOP/s) and maintained a well-balanced workload across the nodes, utilizing a total of 4018.27 GB (4.02 TB) of memory. Despite the high memory usage, no significant bottlenecks were observed in the network, indicating that the MPI communication between nodes was efficient. Fig.~\ref{MPI_Communication_and_Memory_Usage_SST} highlights the breakdown of MPI communication functions, revealing critical results: 
\begin{itemize}
    \item \textbf{MPI\_Recv} (30.30\%), handles receiving data from other nodes, ensuring that necessary data are available for computation.
    \item \textbf{MPI\_Gatherv} (21.30\%), aggregates data from all nodes, playing a crucial role in collecting results across the distributed system.
    \item \textbf{MPI\_Bcast} (13.49\%), distributes data to all nodes, ensuring consistency throughout the simulation.
    \item \textbf{MPI\_Allreduce} (9.19\%), combines data across nodes to perform global reductions, such as sums or averages.
    \item \textbf{MPI\_Wait} (8.49\%), synchronizes MPI processes, ensuring that no process proceeds until others have completed necessary steps.
    \item \textbf{MPI\_Send} (7.68\%), sends data between nodes, facilitating communication across the system.
\end{itemize}
While the hybrid BIT1 openPMD SST simulation demonstrates efficient MPI communication and scalability across nodes, the notable overhead associated with data exchange, aggregation, and synchronization highlights areas that could benefit from further investigation as the system scales.

\begin{figure}[!ht]
   \vspace{0cm} 
    \begin{center}
        \includegraphics[width=\linewidth]{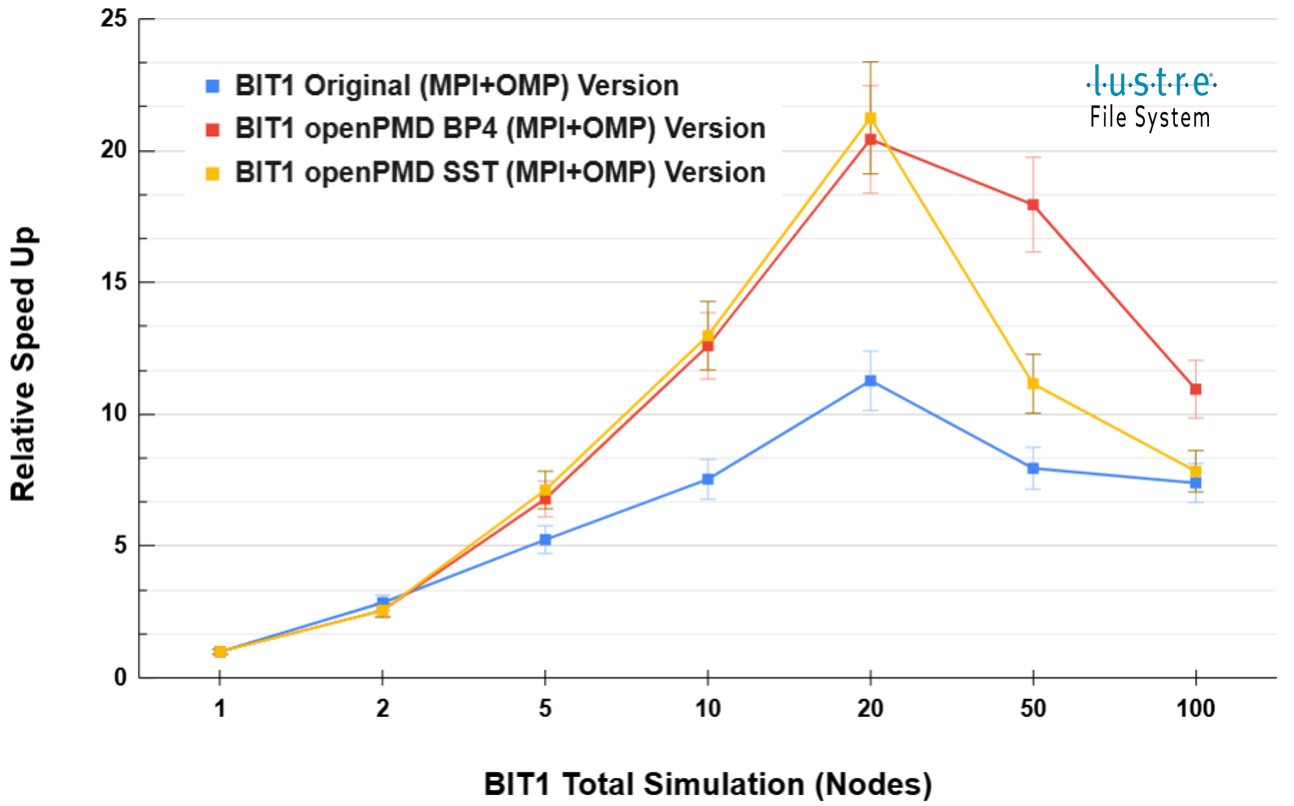}
        \caption{Hybrid BIT1 Total Simulation (Relative) Speed Up - strong scaling up to 100 Nodes (12,800 MPI ranks) on \emph{Dardel} for 200K times steps.} 
        \label{BIT1_openPMD_MPI_OMP_All}
    \end{center}
    \vspace{-0.4cm} 
\end{figure}

Using \texttt{perf}, an open-source profiling tool, we reveal interesting scaling trends of the hybrid BIT1 code across different node configurations. As shown in Fig.~\ref{BIT1_openPMD_MPI_OMP_All}, the relative speedup performance analysis of the BIT1 Total Simulation reveals clear scaling trends and characteristic variability across the three configurations: Original (MPI+OMP), openPMD BP4 (MPI+OMP), and openPMD SST (MPI+OMP). All versions show strong scaling up to 20 nodes, with the openPMD BP4 and SST configurations outperforming the Original version from 5 nodes onward. At 20 nodes, the Original version achieves an 11.27× speedup, while openPMD BP4 and SST reach 20.42× and 21.24×, respectively, with SST achieving the highest overall speedup. Each version peaks at 20 nodes, after which scaling efficiency begins to decline. The increased variability observed beyond this point is reflected by the error bars, which capture runtime fluctuations arising from system-level factors such as communication overhead, I/O contention, and dynamic load imbalance across MPI ranks. The Original version drops to 7.39× speedup at 100 nodes, indicating growing parallel overheads and reduced efficiency. openPMD BP4 decreases slightly at 50 nodes to 17.95× but maintains a relatively strong 10.95× speedup at 100 nodes. openPMD SST experiences a more pronounced decline, falling to 11.16× at 50 nodes and 7.84× at 100 nodes. At 100 nodes, the parallel efficiency drops to 7.4\% for the Original version, 10.95\% for BP4, and 7.84\% for SST, relative to ideal linear scaling. This behavior indicates that all versions face strong-scaling inefficiencies and growing variability beyond 20 nodes, with openPMD SST demonstrating the best peak scaling, while openPMD BP4 provides the most stable sustained performance as the node count increases.

\begin{figure*}[!ht]
\vspace{0cm} 
    \begin{center}
        \includegraphics[width=\linewidth]{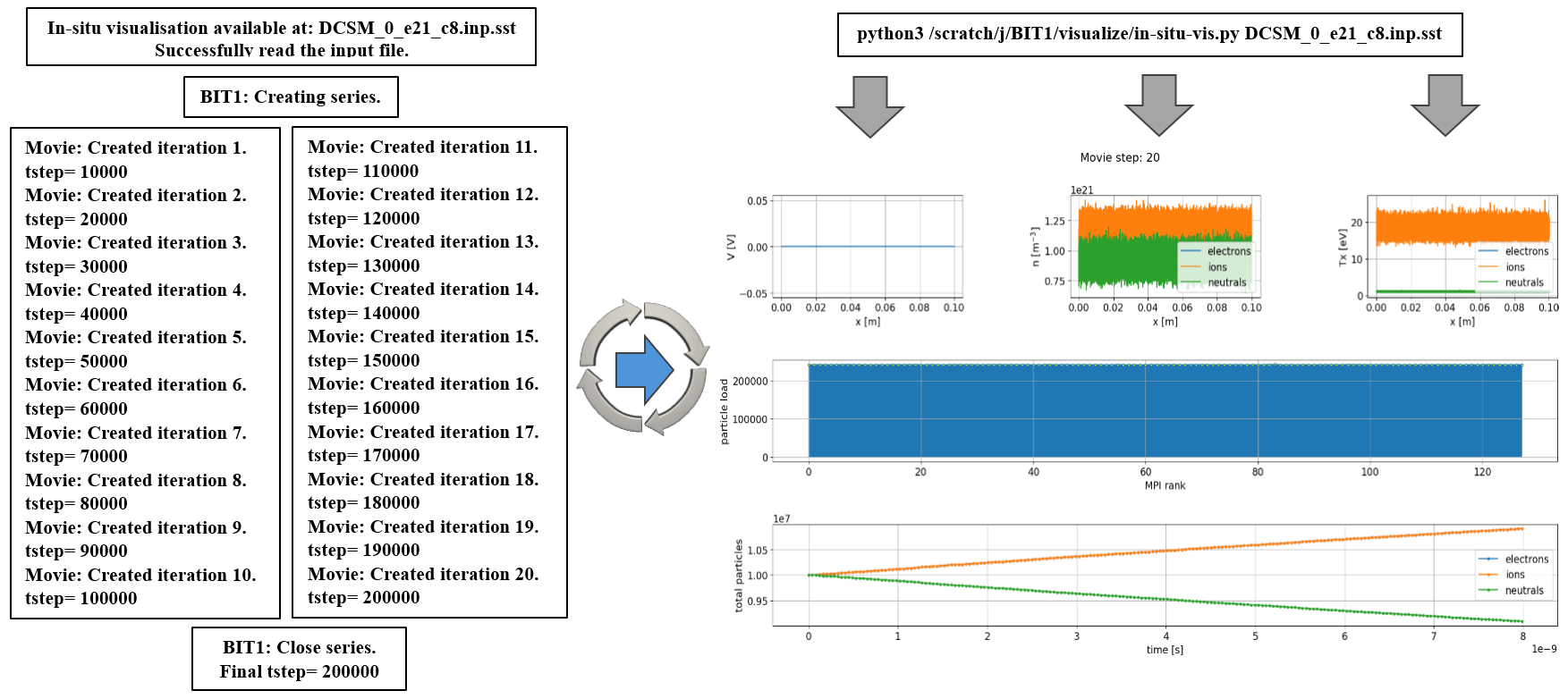}
        \caption{Performing real-time data checkpoint analysis and visualization using the hybrid BIT1 openPMD SST simulation on \emph{Dardel}, up to 200K time steps (tsteps), highlighting a reduced set of plasma profiles, the particle load per MPI rank, and the time evolution of the total number of particles for each species.} 
        \label{BIT1_openPMD_In-situ_Visualizations_Python}
    \end{center}
    \vspace{-0.6cm} 
\end{figure*}

\subsection{Hybrid BIT1 openPMD SST In-Situ Visualization}
One of the key aspects of this work is enabling real-time data checkpoint analysis and visualization in large-scale PIC MC simulations without interrupting the simulation. This addresses I/O bottlenecks and minimizes data movement overhead. Traditional post-processing, as shown in Fig.~\ref{Post_Processing_BIT1}, requires storing vast amounts of data, which can significantly impact performance. In contrast, real-time data checkpoint analysis allows data to be processed and visualized directly within the simulation, eliminating the need for extensive storage and data transfer.

In this work, hybrid BIT1 leverages openPMD with ADIOS2 SST for low-latency memory-to-memory data streaming, enabling seamless real-time data checkpoint analysis and visualization. Unlike BP4, which writes data to disk for later processing, SST allows for immediate interaction with live simulation data, enabling continuous analysis. Fig.~\ref{BIT1_openPMD_In-situ_Visualizations_Python} illustrates the information displayed to a BIT1 HPC system user on Dardel compute nodes after logging in, showing the hybrid BIT1 openPMD SST simulation performing real-time data checkpoint analysis and visualization. A customized Python script, utilizing openPMD’s streaming API and the ADIOS2 SST engine, facilitates the process. Visualizations include:
\begin{itemize}
    \item \textbf{On the first row} (from left to right), a reduced set of plasma profiles (i.e., physical quantity vs position) is plotted: 
        \begin{itemize}
            \item \textbf{electric potential} $(x\,[\mathrm{m}],\, V\,[\mathrm{V}])$: to see the presence of plasma sheath (if applicable).
            \item \textbf{densities of plasma species} $(x\,[\mathrm{m}],\, n\,[\mathrm{m^{-3}}])$: to monitor the particle transport.
            \item \textbf{temperatures of plasma species} $(x\,[\mathrm{m}],\, Tx\,[\mathrm{eV}])$: to monitor the heat transport.
        \end{itemize}
    \item \textbf{On the second row} $(\text{MPI rank},\, \text{particle load})$, particle load per MPI rank is plotted to visualize how much work each MPI rank is doing. If some MPI ranks are overwhelmed compared to the rest of them, then a restart with load balancing (\texttt{--streaming} and \texttt{--load\_balance} parameters) can be initialized to spread the work evenly among the MPI ranks.
    \item \textbf{On the third row} $(time\,[\mathrm{s}],\, \text{total particles})$, the time evolution of total particles of each species is plotted. This information is needed in order to assess if the simulation has reached a steady state or not. If it did, this means that the plasma sources and sinks reached a balance that the number of all species does not change over time anymore. 
\end{itemize}
In the ``Neutral Particle Ionization (Hybrid BIT1) Simulation", where uniform plasma interacts with deuterium gas, causing ionization, and as shown in Fig.~\ref{BIT1_openPMD_In-situ_Visualizations_Python}, electron and ion numbers increase due to neutral ionization, while neutral particles decrease over time. In-situ analysis and visualization provide immediate insights into system states, enabling prompt adjustments to simulation parameters without interrupting execution.

\section{Related Work}
BIT1, an advanced PIC MC code for simulating plasma-material interactions, is extensively used in fusion research, including tokamak simulations~\cite{tskhakaya2010pic}. It builds on the XPDP1 code~\cite{verboncoeur1993simultaneous}, developed by Verboncoeur’s team at Berkeley, and incorporates an optimized data layout for efficient handling of collisions~\cite{tskhakaya2007optimization}. Chaudhury et al.~\cite{chaudhury2019hybrid} applied hybrid parallelization to low-temperature plasma simulations, using OpenMP and MPI for scalable performance on HPC clusters. Recently, Williams et al.~\cite{williams2023leveraging,williams2024understandingp2} identified the possibility to optimize particle mover using OpenMP task-based parallelism~\cite{williams2024optimizing}. In a separate study, Williams et al.~\cite{williams2024enabling} tackled parallel I/O challenges by incorporating openPMD with BIT1 and using Darshan for I/O monitoring. Poeschel et al.~\cite{poeschel2021transitioning} shifted from traditional file-based workflows to streaming data pipelines with openPMD and ADIOS2, highlighting the benefits of streaming I/O for better flexibility and performance in HPC environments. Huebl et al.~\cite{openPMDstandard,openPMDapi} developed the openPMD-api, a C++ and Python API to standardize I/O and facilitate integration with in-situ analysis tools, promoting interoperability across simulation codes. Williams et al.~\cite{williams2024understanding} investigated the impact of openPMD on BIT1’s performance through instrumentation, monitoring, and in-situ analysis. Williams et al.~\cite{williams2024characterizing} also characterized the performance of the implicit massively parallel PIC code iPIC3D, analyzing communication efficiency, optimal node placement, overlapping communication with computation, and load balancing to improve large-scale 3D plasma simulations relevant to magnetic reconnection and space physics, with profiling and tracing guiding recommended communication strategies for advancing Geospace Environmental Modeling challenges. As in-situ analysis becomes crucial for large-scale simulations, Ayachit et al.~\cite{ayachit2015paraview} introduced ParaView Catalyst to integrate data processing into the simulation workflow, reducing storage needs and improving real-time insights. Similarly, Karimabadi et al.~\cite{karimabadi2013situ} explored in-situ visualization for global hybrid simulations, demonstrating its advantages in reducing data movement and enhancing temporal resolution for scientific analysis. Building on these developments, libraries such as Fides~\cite{pugmire2021fides} provide flexible, high-performance data models for both post hoc and in-situ visualization, enabling efficient streaming and zero-copy access to complex simulation data.


\begin{figure*}[!ht]
\vspace{0cm} 
    \begin{center}
        \includegraphics[width=\linewidth]{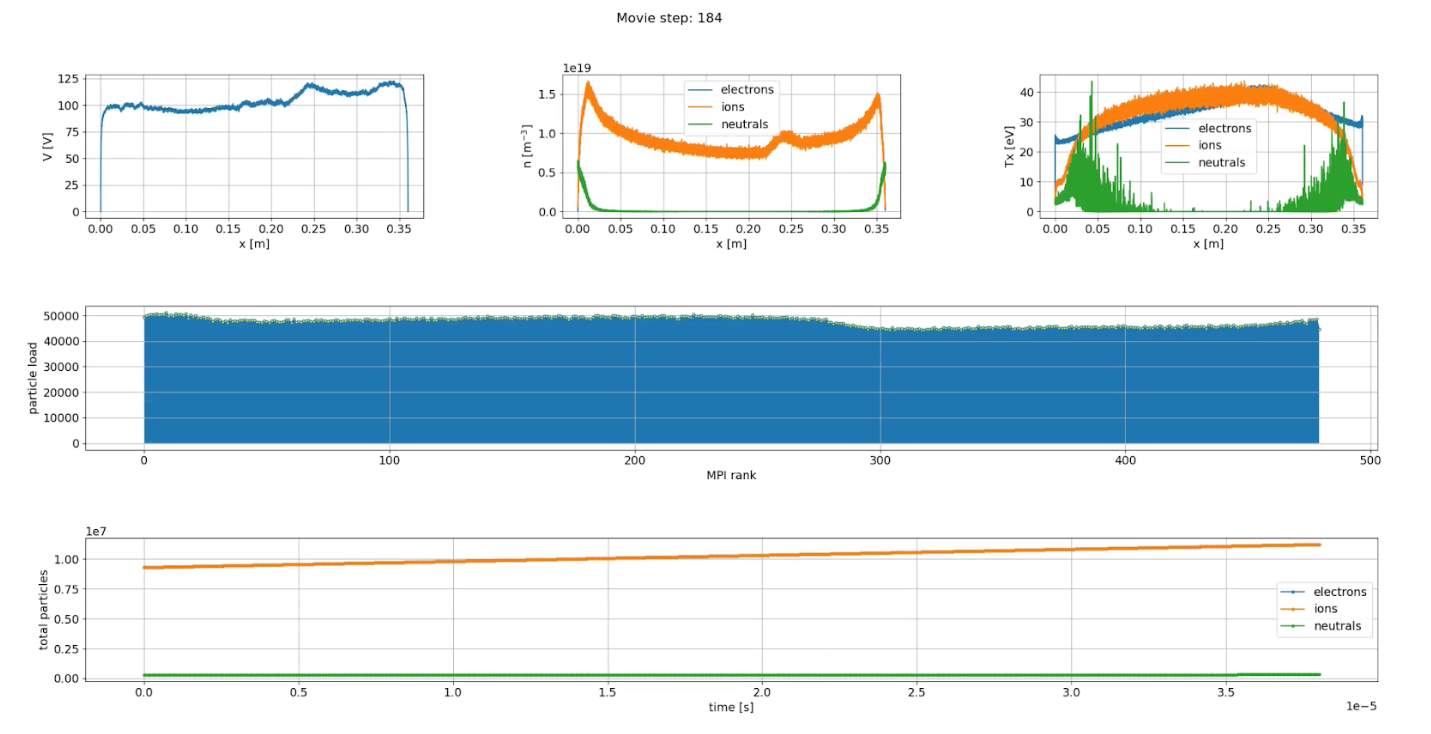}
        \caption{Performing real-time checkpointing, data analysis, and visualization of a magnetized plasma tube bounded by grounded walls.} 
        \label{BIT1_Magnetized_Plasma_Tube}
    \end{center}
     \vspace{-0.4cm} 
\end{figure*}

\section{Discussion \& Future Work}
This work presents the integration of openPMD with ADIOS2's SST backend into the hybrid BIT1 code. This novel method has effectively addressed the data management and storage (I/O) challenges, enhancing I/O performance and computational efficiency for the fusion energy community. The adoption of OpenMP task-based parallelism has optimized the particle mover function, improving load balancing and reducing execution times. Profiling tools such as gprof, perf, IPM, and Darshan have provided valuable insights into computation, communication, and I/O operations, confirming the effectiveness of this approach. The transition from traditional file-based I/O to ADIOS2's streaming capabilities has reduced metadata overhead and write times, further improving overall simulation performance.

In-situ visualization, enabled through openPMD's streaming API and the ADIOS2 SST engine, facilitates real-time data analysis without interrupting the simulation workflow. This advancement provides immediate insights into plasma behavior, enabling prompt adjustments to simulation parameters and more efficient use of computational resources. However, as simulations scale to larger node counts, increased communication overhead and synchronization costs have been observed, highlighting areas for further optimization to maintain scalability and performance. It is important to note that this analysis was conducted for a moderate-size problem (neutral particle ionization case), which is not representative of large-scale production workloads. As such, better scalability is expected for the largest production cases, where I/O can be effectively \emph{hidden} behind the now computation-intensive workload, up to the hardware bandwidth limit of the machine. Visualizations can be of the form shown in Fig.~\ref{BIT1_Magnetized_Plasma_Tube}, performing real-time checkpointing, data analysis, and visualization of a magnetized plasma tube bounded by grounded walls.

\begin{figure*}[!ht]
\vspace{0cm} 
    \begin{center}
        \includegraphics[width=\linewidth]{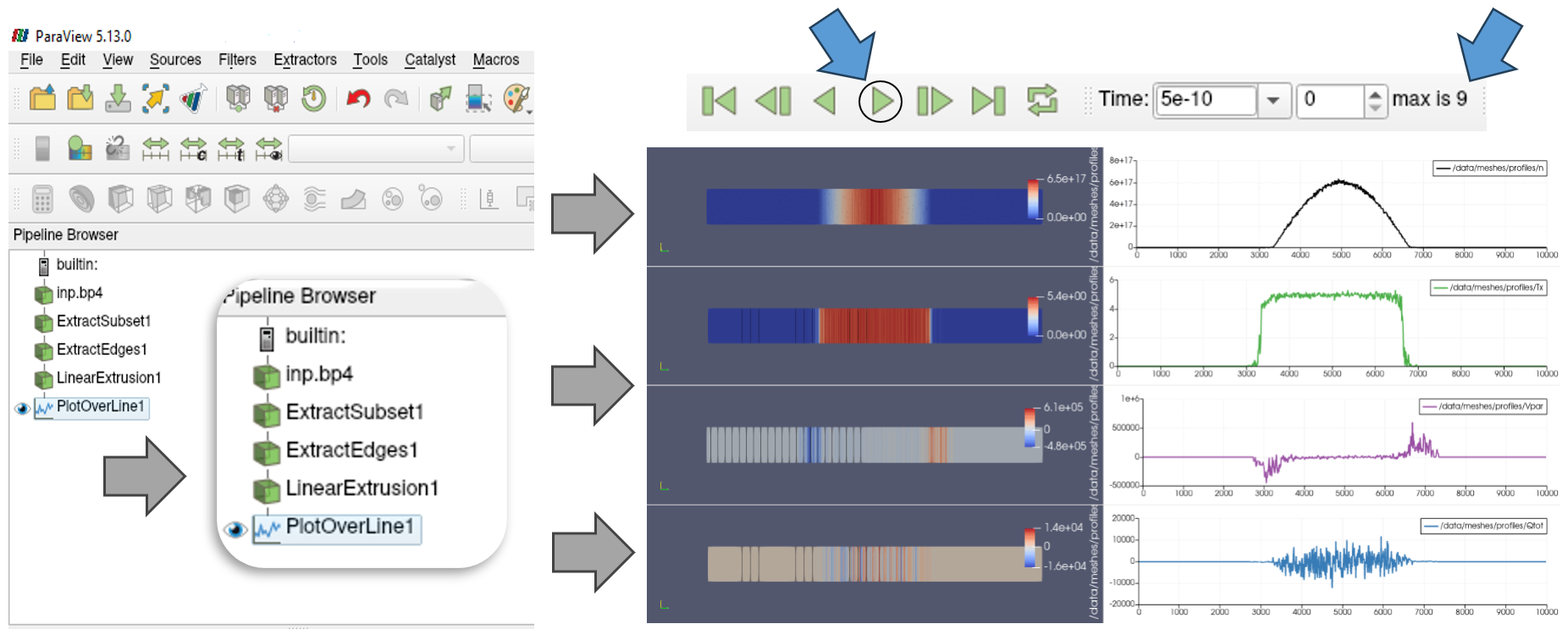}
        \caption{Hybrid BIT1 openPMD BP4, on \emph{Dardel}, using the \texttt{--append} parameter to load all checkpoints (0 - 9) at once in ParaView to plot and visualize (blue arrows) the evolution of profiles. The left plots show 3D representations of 1D data with color-coded physical quantities, while the right plots display the same 1D data in 2D representations.} 
        \label{BIT1_openPMD_In-situ_Visulaization_Paraview}
    \end{center}
     \vspace{-0.4cm} 
\end{figure*}

Future research will focus on enhancing hybrid BIT1 simulations by leveraging the strengths of the BP4 and SST backends. BP4 is optimized for high-throughput storage through efficient indexing and compression, while SST enables low-latency, memory-to-memory data streaming for real-time analysis and visualization. The integration of openPMD and ADIOS2 with ParaView further improves the efficiency of data visualization and analysis. ADIOS2 supports high-performance I/O by using BP4 for checkpoint storage and SST for real-time data streaming. Hybrid BIT1 outputs are stored in multidimensional arrays, typically indexed first by particle species. Within ParaView, filters process these arrays to extract and visualize key physical quantities such as density, temperature, velocity, and energy flux for selected species. As shown in Fig.~\ref{BIT1_openPMD_In-situ_Visulaization_Paraview}, the \texttt{inp.bp4} file represents the data stream produced by the load file function, in this case handled by the \texttt{ADIOS2CoreImageReader}. The \texttt{ExtractSubset1} filter is used to isolate a per species one-dimensional array containing the desired physical quantity from the initial multidimensional dataset. This is followed by the \texttt{ExtractEdges1} filter, which interpolates cell data onto mesh edges, a critical step before applying the \texttt{LinearExtrusion1} filter. The \texttt{LinearExtrusion1} filter then extrudes the 1D data into 2D, making it more suitable for visualization in ParaView’s 3D environment. Finally, \texttt{PlotOverLine1} reproduces the graphs generated by the original Python script, enabling consistent analysis directly within the ParaView interface. The plots on the left show a 3D representation of 1D data, where the second dimension is created by extending the first. The values of the physical quantity of interest are color-coded within the volume of the 3D visualization, and the color-value relationship is indicated by the color bar in each plot. The plots on the right show 2D representations of the same 1D data, where the values are displayed along the second dimension. Using ParaView instead of a customized Python script for data visualization can allow hybrid BIT1 users to load all checkpoints at once and visualize the evolution of profiles over time. ParaView also provides flexible options for adjusting plot colors and styles without requiring code modifications.

The integration of openPMD and Catalyst-ADIOS2 with ParaView~\cite{mazen2023catalyst} will also be explored. This approach enables data reduction on simulation nodes using Catalyst, followed by the transfer of reduced data to dedicated visualization nodes for in-transit processing. These hybrid analysis frameworks aim to accelerate scientific discovery by supporting detailed real-time visualization and optimizing both post-processing and in-situ analysis for large-scale PIC MC simulations.

\begin{acks}
Funded by the European Union. This work has received funding from the European High Performance Computing Joint Undertaking (JU) and Sweden, Finland, Germany, Greece, France, Slovenia, Spain, and Czech Republic under grant agreement No 101093261 (Plasma-PEPSC). The computations/data handling were/was enabled by resources provided by the National Academic Infrastructure for Supercomputing in Sweden (NAISS), partially funded by the Swedish Research Council through grant agreement no. 2022-06725.
\end{acks}


\bibliographystyle{sagev}

\end{document}